\documentclass[galaxies,review,accept,oneauthor,pdftex]{Definitions/mdpi}

\firstpage{1}
\pubvolume{xx}
\issuenum{1}
\articlenumber{5}
\pubyear{2019}
\copyrightyear{2019}
\history{Received: 30 January 2019; Accepted: 23 April 2019; Published: date}
\updates{yes} 

\setitemize{parsep=6pt,itemsep=0pt,leftmargin=*,labelsep=5.5mm}
\setenumerate{parsep=6pt,itemsep=0pt,leftmargin=*,labelsep=5.5mm}
\setlist[description]{itemsep=0mm}

\usepackage{array}
\usepackage{graphicx}
\usepackage{amssymb}

\definecolor{red}{rgb}{1,0,0}

\newcommand{\hlo}[1]{#1}
\renewcommand{\hl}[1]{}

\newcommand{\wmap}{\textit{WMAP}}
\newcommand{\planck}{\textit{Planck}}

\newcommand*\aap{A\&A}
\newcommand*\aanda{A\&A}

\newcommand*\apj{ApJ}

\newcommand*\apjs{ApJS}

\Title{Practical Modeling of Large-Scale Galactic Magnetic Fields: Status and Prospects}

\Author{Tess~R. Jaffe $^{1,2}$}

\AuthorNames{Tess~R. Jaffe}

\address{%
$^{1}$ \quad Department of Astronomy, University of Maryland, College Park, MD 20742, USA\\
$^{2}$ \quad CRESST II, NASA Goddard Space Flight Center, Greenbelt, MD 20771, USA\hlo{; tess.jaffe@nasa.gov}
}

\abstract{This is a review of the status of efforts to model the large-scale  Galactic magnetic field (GMF).  Though important for a variety of astrophysical processes, the GMF remains poorly understood despite some interesting new tracers being used in the field.  Though we still have too many models that might fit the data, this is not to say that the field has not developed in the last few years.  In~particular, surveys of polarized dust have given us a new observable that is complementary to the more traditional radio tracers, and a variety of other new tracers and related measurements are becoming available to improve current modeling.  This paper reviews:  the~tracers available;  the~models that have been studied;  what has been learned so far;  what the caveats and outstanding issues are;  and one opinion of where the most promising future avenues of exploration lie.
}


\begin{document}

\section{Introduction}

The morphology of the large-scale Galactic magnetic field (GMF) is surprisingly poorly understood for such an important component of the Milky Way's interstellar medium (ISM).  There is a long list of topics in Galactic astrophysics that currently depend on an incomplete understanding of the GMF, such as disk dynamics, cosmic-ray propagation, the turbulent ISM, molecular cloud collapse, star formation, supernova remnant evolution, etc.  There is also a list of studies in the literature modeling the GMF using a variety of observational tracers and parametrized morphological forms.  For earlier reviews, see, e.g., \cite{haverkorn:2014,Beck:2003ke} and references therein.  In this review, I will discuss modeling work that is either relatively recent or still being used

In addition to its importance in its own right, the GMF has a significant impact on several extragalactic observations because of effects it can have on the observables or because of the foreground confusion it adds. The~cosmic microwave background (CMB) is the most obvious example, since we observe the CMB in the foreground minimum between the synchrotron emission that dominates in the radio (Section~\ref{sec:obs_sync}) and the dust emission that dominates the higher frequencies (Section~\ref{sec:obs_dust}).
%
%
Another~cosmological problem impacted by the Galaxy foregrounds is the study of the recombination epoch using redshifted 21\,cm emission, which may be contaminated by Galactic synchrotron emission that is orders of magnitude brighter (\cite{Zaroubi:2012kt} Section 5.5).  The~search for the sources of the highest energy particles in the Universe, ultra-high-energy cosmic rays (UHECRs) is also complicated by the fact that these particles are deflected by magnetic fields as they propagate to the Earth, so back-tracing them requires an accurate GMF model (see Section~\ref{sec:crs}).  These needs have driven some of the modeling work in the field and will continue to do so.

The variety of observables and the variety of contexts where the GMF is important explain the variety of modeling efforts in the literature.  Some of these analyses focus on only one part of the problem and, as in the case of the proverbial elephant, find an answer that is incomplete.  All~of the analyses require certain assumptions about the distributions of particles associated with the observables, whether thermal electrons, relativistic cosmic rays, or dust grains.  These assumptions mean that there remain hidden degeneracies in the parameter space and that different analyses come up with different measures even for fundamental quantities such as the average strength or degree of ordering of the field.  However, the information we can gain from these disparate efforts can also help us to tackle the problem from different angles.

I will summarize the observables in Section~\ref{sec:observables} and the components of a physical model of the Galaxy needed to simulate them in Section~\ref{sec:phys_components}.   I will review some of the different models, their origins and contexts, their advantages and disadvantages, what common features they have,  and what we have learned in Section~\ref{sec:models}.  There are a variety of challenges faced by all such modeling efforts, which I will summarize in Section~\ref{sec:challenges}.  However, there is also a prospect of making significant progress in the next few years, which I will speculate on in Section~\ref{sec:prospects}.

\section{Observables}\label{sec:observables}

There are many physical processes affected by the GMF that allow us to observe it indirectly.  None of these phenomena give complete and unambiguous information about the large-scale GMF by themselves.  This section will briefly summarize the principle phenomena that have been used so far to model the Galactic-scale GMF, with the analyses and results summarized in the next section.  A~summary of this section is also presented in Table\,\ref{tab:observables}.

There are other tracers not discussed here, from Zeeman splitting of masers (\cite{Fish:2003gp,Han:2006hs,Green:2012fm}) to HI velocity gradients (\cite{GonzalezCasanova:2017fh}).  For a thorough review of observations and analyses from small-scale turbulence to the intergalactic medium (IGM), see \citet{han:2017}.  Here we will focus on probes that probe the large-scale magnetic field over large portions of the Galaxy. 

{ 

\begin{table}[H]
\caption{Table of large-scale GMF tracers, their pros, cons, and dependencies.} \label{tab:observables}
\centering

\begin{tabular}{>{\raggedright\arraybackslash}m{2.cm}>{\raggedright\arraybackslash}m{2.5cm}>{\raggedright\arraybackslash}m{3cm}>{\raggedright\arraybackslash}m{3cm}>{\raggedright\arraybackslash}m{3cm}}
\toprule
\textbf{Observable}  & \textbf{GMF Property Probed} & \textbf{Dependencies}     & \textbf{Pros}      & \textbf{Cons}  \\ \midrule

Starlight polarization & $B_\perp$ orientation        & dust grain properties and distribution                  & 3D information  & sampling limited to a few kpc  \\\midrule

Faraday rotation (extragalactic)  & $B_\parallel$ direction and strength  &  thermal electron density    & good full-sky sampling (42k~sources);\linebreak full LOS through Galaxy  & no 3D info along LOS \\\midrule

Faraday rotation (Galactic)  & $B_\parallel$ direction and strength  & thermal electron density    & 3D sampling along the LOS through the Galaxy  & mostly in Galactic plane;\linebreak currently insufficient sampling ($1$k) \\\midrule

Faraday tomography (extragalactic) & $B_\parallel$ direction and strength &  thermal electron density   & probes variations along the LOS through the Galaxy & low physical resolution, not a probe of the Milky Way \\\midrule

Faraday tomography (Galactic) & $B_\parallel$ direction and strength  &  thermal electron density   & probes variations in 3D along the LOS & no physical distances associated with Faraday depth variations\\\midrule

Diffuse synchrotron emission\linebreak (radio) & $B_\perp$ orientation and strength (squared) & cosmic-ray density;\linebreak thermal electron density & goes as $|\mathbf{B}|^2$;\linebreak full-sky coverage;\linebreak probes turbulent Faraday effects & no 3D info along LOS;\linebreak polarization horizon of a few kpc due to Faraday depolarization effects  \\ \midrule

%


Diffuse synchrotron emission \linebreak (microwave) & $B_\perp$ orientation and strength (squared) & cosmic-ray density & goes as $|\mathbf{B}|^2$;\linebreak full-sky coverage;\linebreak full LOS through the Galaxy;\linebreak no Faraday rotation & no 3D info along LOS;\linebreak total intensity contaminated by Bremsstrahlung and AME.  \\ \midrule

Diffuse dust emission & $B_\perp$ orientation & dust grain density, properties, environment, and alignment & full-sky coverage;\linebreak full LOS through the Galaxy;\linebreak 3D information with extinction surveys (e.g., Gaia)\linebreak no Faraday rotation & probes only close to Galactic plane $|z|\lesssim 100$\,pc  \\

\bottomrule
\end{tabular}
\end{table}

} 

\subsection{Polarized Starlight}

One of the longest-known observational signatures of the GMF is the polarization of starlight.  Amorphous dust grains tend to align their long axes perpendicular to the local magnetic field, and therefore absorption leaves the starlight partly linearly polarized parallel to that local field as projected onto the sky from the point of view of the observer (perpendicular to the line of sight, i.e., $B_\perp$).

The catalog of \citet{heiles:2000}, for example, provides measurements to over 9k individual stars over a significant fraction of the sky.  The~advantage of starlight polarization is that one can in principle use the distance to the star as well as multiple stars in a given direction to extract 3D information about the magnetic field.  This analysis depends on knowledge of the distribution of the relevant dust grains, of course, which can be measured via the reddening.  Its main disadvantage for modeling the large-scale GMF is sampling, since we cannot make such measurements, much less get accurate distances, for stars in the more distant regions of the disk.  We also run out of stars away from the Galactic plane.  However, \citet{santos:2011} demonstrate how useful they can be for studying local features, particularly the North Polar Spur (NPS, aka Loop I), and \citet{Pavel:2011fn} explore how near infra-red observations can be used to constrain large-scale field properties.
See Section~\ref{sec:local_features}.  \citet{Panopoulou:2018uh} demonstrate how combining polarization measurements with Gaia\footnote{\url{https://www.cosmos.esa.int/web/gaia}.} distances allows a detailed tomography of the~ISM.

\subsection{Faraday Rotation Measures (RMs) of Point Sources}

The Faraday rotation of polarized emission can probe the line-of-sight (LOS) component of the magnetic field ($B_\parallel$) through the wavelength-dependent rotation of the orientation of the linear polarization vector of emission originating from a source behind a Faraday rotating medium.  For~each polarized point source observed at multiple frequency bands, a~single RM value can be fit to the polarization orientation as a function of frequency, and this RM represents the integrated product of the thermal electron density and the LOS component of the magnetic field between the observer and the source.  This observable is unique in that it probes not only the orientation of the magnetic field but also its direction:  a positive (negative) RM implies a field pointed toward (away from) the observer.  RMs of Galactic pulsars can give us 3D information if we have a measured distance to each pulsar, but our sampling is currently very limited.  RMs of extragalactic point sources are far more plentiful, but~each measurement represents the full LOS through the Galaxy (as well as the intergalactic medium) in that direction as well as the Faraday rotation intrinsic to the source itself.

We currently have roughly 1k RMs for Galactic pulsars \cite{han:2018} and ~42k for extragalactic polarized sources \cite{taylor:2009, Xu:2014ic} and more all the time \cite{Schnitzeler:2019ge}.  The~sampling of the Galactic pulsars allows some analysis of the field morphology (see, e.g., work by \citet{han:2018}) but is not currently sufficient to use for robust model fitting and to take full advantage of the 3D information this tracer has the potential to offer.  (That situation will change, however, with the Square Kilometer Array (SKA)\footnote{\url{https://www.skatelescope.org/}.} and its associated pathfinder surveys;  see Section~\ref{sec:prospects}) Extragalactic sources, however, are plentiful over the full sky and particularly well sampled over the Southern and Canadian Galactic Plan Surveys (SGPS and CGPS \cite{brown:2003,brown:2007}.  Many catalogs of extragalactic RMs have been collected by \citet{oppermann:2012b} and synthesized into a reconstructed sky map of observed total RM through the Galaxy as well as an uncertainty based on the sampling and the variations among the data used in any given region of the sky.

\subsection{Diffuse Polarized Synchrotron Emission}\label{sec:obs_sync}

Diffuse synchrotron emission dominates the sky in maps from the radio frequencies to the microwave bands.  It depends both on the strength and orientation of the magnetic field projected onto the plane of the sky ($B_\perp$).  One of the first full-sky maps and one that remains useful today is that of Haslam et al. at 408\,MHz \cite{haslam:1982, remazeilles:2015} giving total synchrotron intensity, $I$, from a combination of ground-based single-dish data.  Polarization (in the form of Stokes parameters $Q$ and $U$, or polarized intensity as $PI\equiv\sqrt{Q^2+U^2}$) has been measured over the full sky in the radio at, for example, 1.4\,GHz by \citet{reich:2001b} and \citet{testori:2008}, showing large-scale coherent polarization signals in the NPS and Fan regions as well as significant depolarization in the Galactic plane compared to the high latitude sky.  This is the main disadvantage of probing the polarization at radio frequencies, where it is easier to do from the ground:  there is a so-called polarization horizon \cite{uyaniker:2003,Hill:2018gc}, which is a function of telescope beam size and observation frequency, beyond which little polarization signal can be observed due to Faraday effects in the turbulent ISM.

Space-based microwave observations began with \wmap\ that showed us the 23\,GHz sky in polarized emission free of Faraday effects \citep{wmap_9yr}, and similarly at 30\,GHz by \planck\  \citep{planck:2018}.  These data allow us to study the apparent morphology of the magnetic fields projected onto the sky.  One of the most important observables, however, that of the polarization fraction, $p\equiv PI/I$, remains unavailable to us.  This is because the total intensity sky at microwave bands is dominated by other emission processes, principally thermal Bremsstrahlung emission and the anomalous microwave emission (AME) believed to arise from spinning dust grains.  Both processes are thought to produce only unpolarized emission, but their presence makes it difficult to map the synchrotron total intensity in the microwave bands and therefore to estimate the degree of polarization and the field ordering.  An additional complexity is the calibration of the zero-level of these maps;  see \citet{wehus:2014} for a recent analysis.  Most of these radio and microwave observations are not absolutely calibrated, which means that there is an unknown net offset in the datasets, and though this offset is not important for the fitting of morphological features, it is again vital for the polarization fraction and inferred field ordering.

If we understood the energy spectrum of the cosmic-ray lepton population that produces the synchrotron emission from the radio to the microwave bands, we could combine the low-frequency total intensity maps (where contamination is minimal) with the high-frequency polarization maps (where Faraday effects are minimal) to measure the polarization fraction.  Unfortunately, the shape of the spectrum and its likely turnover around a few GeV are not well understood.  Since direct measures of the cosmic-ray electron (CRE) spectrum near the Earth are additionally complicated by solar modulation, and the local spectrum may not be typical of the Galaxy, the synchrotron emission itself may be the best indirect probe of that region of the CRE spectrum \citep{jaffe:2011} if we can combine information from enough different frequencies around a few GHz.  This situation is improving because of the C-Band All Sky Survey (C-BASS) at 5\,GHz \citep{king:2014,irfan:2015bm}, precisely the region most interesting for probing not only the synchrotron spectral turnover but also the regime where Faraday effects go from dominant to negligible in different regions of the plane.

\subsection{Diffuse Polarized Thermal Dust Emission}\label{sec:obs_dust}

The same dust grains that polarize background starlight through absorption also produce thermal emission that is polarized perpendicular to the local magnetic field.  The~observed dust polarization is then another tracer of the orientation of the magnetic field projected onto the sky.  It is not thought to be a strong function of the magnetic field strength, but the degree to which the grains tend to align depends on the grain properties and environment in ways  not well understood.  This was first measured by Archeops \cite{Benoit:2004} and more recently with the full-sky high-resolution and multi-frequency data of the \planck\ mission \cite{pipXIX, planck:2018}.

Though the geometric dependence is similar to that of synchrotron emission (i.e., polarization perpendicular to the $B_\perp$ orientation), this observable is complementary to the synchrotron emission, because it arises from a different region of the ISM, the cold dusty ISM close to the Galactic plane.  The~\planck\ data showed \citep{pipXIX} that though the two tracers correlated in some regions such as the Fan and NPS, they were uncorrelated over much of the sky.  This means that we can use the different particle distributions to probe the field along the LOS.  Though detailed models of the distribution of CREs are lacking, the Gaia mission is providing new 3D dust models from the extinction measurements of millions of stars \cite{Lallement:2018kl}.  The~combination of Gaia and \planck\ data will allow us to probe the local magnetic field in 3D using the dust emission.  This in turn will then help us to determine which aspects of the large angular-scale structure in the synchrotron sky are local (see Section~\ref{sec:local_features}) and let us account for them.  This in turn would then let us focus on the vertical structure of both dust and synchrotron emission to probe from the disk into the Galactic halo out to a few kpc.

\subsection{Diffuse $\gamma$-ray Emission}

Diffuse $\gamma$-ray emission has long been used to study the cosmic-ray population in the Galaxy and is therefore crucial to interpreting the synchrotron emission and studying the GMF.  The~Fermi\footnote{\url{https://fermi.gsfc.nasa.gov/}.} data in particular provide both direct measurements of the local population of CRs as well as the diffuse inverse Compton $\gamma$-ray emission mapped over the full sky.  Both have been used to constrain the CR population \cite{strong:2010}, to probe the cosmic-ray spectrum by combining $\gamma$-ray and microwave band observables \cite{strong:2011}, and to fit models of the magnetic field and measure the scale height of the CRE population \cite{orlando:2013}.  It is now being recognized that the question of CR propagation cannot be solved without the combination of $\gamma$-ray and synchrotron emission, as demonstrated in recent work by \citet{Orlando:2019tb}.

\subsection{Faraday Tomography/RM Synthesis}

With enough spectral resolution, one can do better than fit a single RM value to the variation of the polarization angle with frequency in a given direction.  A full Fourier analysis converts the emission as a function of frequency into a  measure of the polarized intensity along the LOS as a function of Faraday depth.  This allows us to study diffuse emission where the synchrotron-emitting regions and the Faraday rotating regions are mixed.  The~Faraday depth, though it is not a physical distance scale, provides 3D information about the distribution of emitting regions along the LOS.  It therefore probes the magnetic fields not only through their Faraday effects but also from the synchrotron emission itself.  This sort of analysis avoids the sampling restrictions of pulsar RMs, though in return, it links the Faraday information to the cosmic-ray density distribution.

See \citet{Ferriere:2016kt} for a brief review of Faraday tomography and its prospects.

\subsection{Supernova Remnants}

The morphology of supernova remnants (SNRs) can be a complementary probe of the GMF as shown by \citet{west:2016}.  When the SNR expands into the ambient ISM, it will compress the local magnetic field component that is tangent to the shell (perpendicular to the expansion direction).  This will in turn produce synchrotron emission that is strongest where the field is most compressed, implying that the morphology of the remnant in the radio is in part determined by the orientation of the ambient field relative to the line of sight.  Though there are not many SNRs with such regular morphology, \citet{west:2016} showed that one large-scale GMF model predicted a significantly better agreement with the available observations than another model.  This analysis is therefore a useful addition to the toolbox of informative probes of the GMF.

\section{Modeling Components}\label{sec:phys_components}

There are no direct measurements of the GMF that do not depend on other components of the ISM, namely on the spatial and spectral distributions of the particles that are also involved.  With the observables outlined above, we need to model not only the GMF itself but also:  the dust grains that polarize starlight and emit in the submm bands;  the warm and hot ionized gas that Faraday rotates polarized emission that propagates through it;  the relativistic cosmic-ray leptons that emit the synchrotron emission when interacting with the GMF.  We summarize all these components of any GMF modeling effort in this section.

\subsection{Magnetic Field}\label{sec:B_components}

\begin{figure}[H]
\begin{centering}
\includegraphics[width=.9\textwidth]{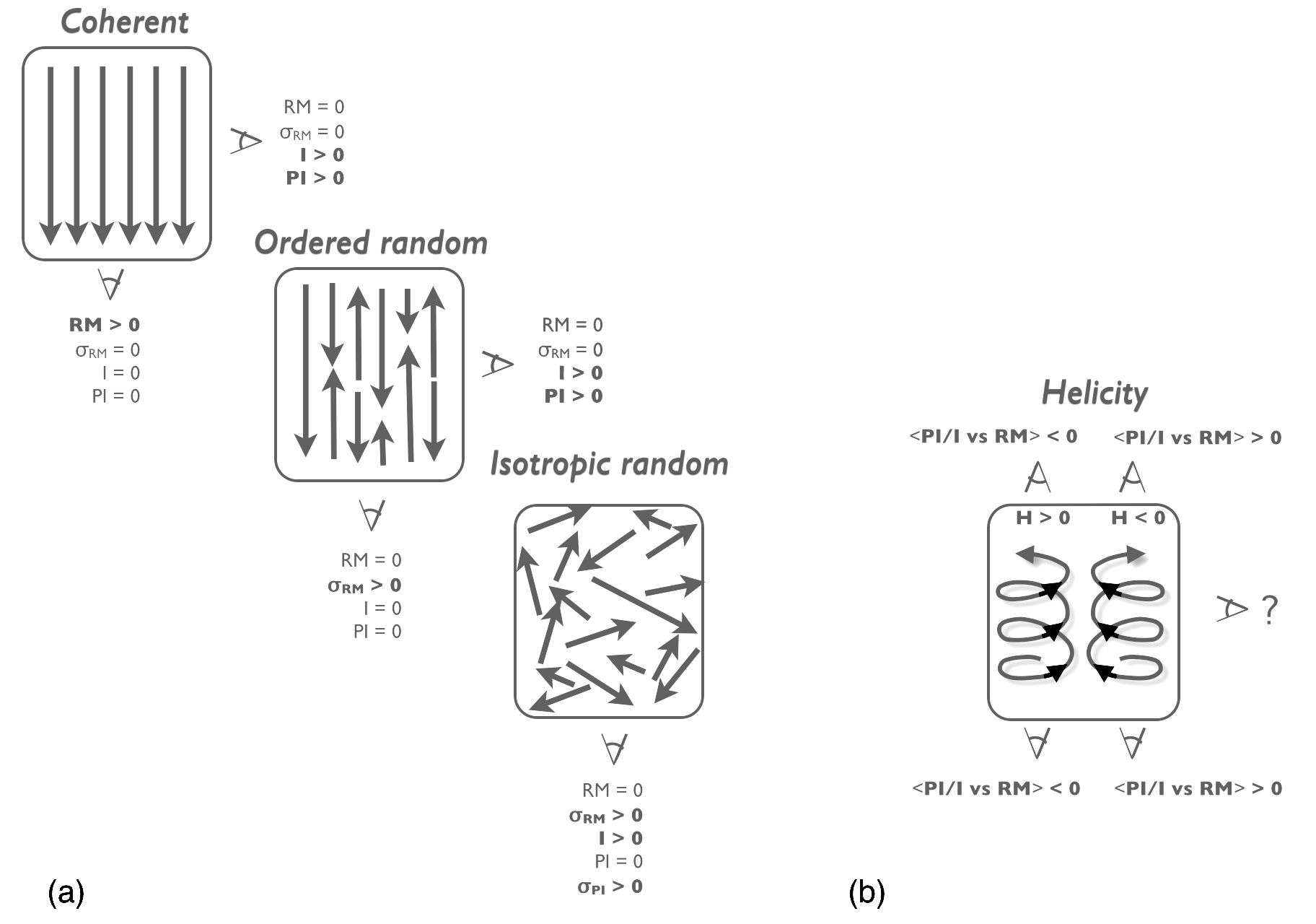}
\end{centering}
\caption{Cartoon illustrating \hlo{several geometrical properties of magnetic fields.  Panel (a) shows} the effective magnetic field components defined by their effects on the indicated observables (from \citet{jaffe:2010} ).  \hlo{Panel (b) shows the helicity.}  See Section~\ref{sec:B_components}.  
\label{fig:cartoon}
}
\end{figure}

Any model for the GMF has several components that are useful to distinguish by their effects on observables and that can be generated separately for computational feasibility.  This~means using ad hoc models rather than full magneto-hydrodynamic (MHD) solutions to the dynamo equations.  This~section summarizes the main characteristics of the GMF that can be probed and by what observables.  A cartoon illustrating these components is shown in Figure\,\ref{fig:cartoon}

\subsubsection{Definitions:  ``Random'', ``Regular'', ``Ordered'', ``Striated''...?}

There is a bit of confusion of terminology in the literature, some of which dates from a time when generally only one observable was studied at a time, and ``regular'' and ``random'' were the only two components of the magnetic field that were discussed.  However, for different observables, these two terms can be ambiguous.  A useful cartoon is shown in Figure\,\ref{fig:cartoon} (originally published by \citet{jaffe:2010}) that demonstrates how the polarized emission  (synchrotron or dust) cannot distinguish the field direction but only its orientation, while the RMs can, and how this combination of observables then divides the field into three effective components.

In brief, the three observables of total synchrotron intensity, polarized synchrotron intensity, and Faraday RMs  divide the GMF into three effective field components:
\begin{itemize}

\item The~coherent field is the component whose {\em direction} remains coherent over large regions.  This is also referred to as the mean field.  When observed from a perpendicular direction, the synchrotron polarization adds coherently, and when observed in parallel, the RM adds coherently.  See left-most box in Figure\,\ref{fig:cartoon}a.

\item The~isotropic random component represents a zeroth-order simplification for the ISM turbulence, where local field fluctuations are equally likely to be in any direction.  Such a component does not have to be single-scale (e.g., as in \citet{sun:2008});  one can define an isotropic Gaussian random field (GRF) that encodes correlations as a function of distance but has equal power in all directions (e.g., as in \citet{jaffe:2010}).  This component contributes only to the total synchrotron intensity but does not add coherently to its polarization or to the RM;  see the lower right box of Figure\,\ref{fig:cartoon}a.

\item The~third component can be thought of representing a first-order approximation of the ISM, where local field fluctuations are not isotropic but rather prefer certain {\em orientations}.  Please note that it does not prefer certain {\em directions};  that would simply be part of the coherent component.  The~definition of this third component is that the RMs must still average to zero but that the polarized emission adds up.  See the middle box of Figure\,\ref{fig:cartoon}a.  This component has variously been called the ``ordered random'' (\citet{jaffe:2010}), or the ``striated'' component (\citet{jansson:2012b}).  The~term ``anisotropic random'' used by \citet{Beck:2003ke} can be ambiguous in that it has sometimes been used to refer to the Figure\,\ref{fig:cartoon}a middle component only and sometimes to the total random component, the sum of the middle and right boxes, i.e., the ordered random plus isotropic random.

\end{itemize}

 It is best to separate these three effective components clearly in the text, and it is imperative to define explicitly what one means if using the term ``regular'' or ``random''.  The~importance of separating these components is in measuring their relative amplitudes and understanding the relationships among them, e.g., how much of the regular field may arise from large-scale differential rotation and shear, or shocks that compress the turbulent component along one direction, etc.  The~next step is then to associate them with the distinct regions or phases of the ISM (see, e.g., \citet{2016arXiv160802398E}) to understand their relationship with the other components of the Galaxy.

\subsubsection{Coherent Field}

The coherent field has been studied with RMs for some time, and the clearest large-scale morphological features are the apparent reversal of the sign of the RMs both across the plane (an~anti-symmetry north to south) and reversals along the Galactic plane over relatively short angular distances ($\sim$10$^\circ$).  The~problem is that the sampling of RMs is not sufficient to resolve unambiguously at what distance these reversals happen and therefore what they imply about the large-scale field morphology.

Regarding the coherent field strength, this in principle can also be estimated from the RMs, but in practice this is limited by our knowledge of the distribution of the thermal electron population and by its correlations with the field .  See discussions by \citet{Beck:2003ke} and \citet{sun:2008}.

\subsubsection{Isotropic Random Field}

The isotropic random field manifests in several ways.  Firstly, as illustrated in Figure\,\ref{fig:cartoon}, this component contributes to the total synchrotron intensity but not to its polarization (or that of dust emission), since the addition of the polarization vectors will average to zero.  This means that the polarization fraction of synchrotron emission could be a good probe of the relative strength of this component.  In practice, this is complicated, as discussed below.  This component can also be probed by measuring the variance of the polarization and RM (e.g., \citet{haverkorn:2004}), since though they average to zero, a~stronger isotropic random component will increase the variance both across the sky and along the LOS.  Again, this is degenerate with variations in the relevant particle distributions and in particular depends on the correlations between the fluctuations in the fields and particles, which are not independent.

\subsubsection{Ordered Random Field}

Without these three complementary observables, this component cannot be disentangled from the others, but it has now been estimated by several teams (\citet{jaffe:2010}, \citet{jansson:2012b}, and \citet{orlando:2013}).  This component is of interest in studying the origin of the turbulence, which is not expected to be isotropic, and the interaction of the small scales with the large-scale dynamics such as shear from differential rotation or compression from spiral arm shocks.  These two mechanisms can generate an ordered but random component from a purely isotropic random component.

\subsubsection{Helicity}

Helicity is a quantity that is recently becoming interesting in studies of magnetic fields, both Galactic and extragalactic, because of its importance to dynamo theory (see, e.g., \cite{Brandenburg:2004kl}).  The~three effective field components discussed above are of course an incomplete description of the GMF;  they are simply defined by the three observables of total synchrotron intensity, polarized synchrotron intensity, and RM.  Helicity is another thing entirely and one that we may be able to probe with either different observables or by looking at these observables differently.

Helicity in a field is not defined at a point but within a volume as
\begin{equation}
H=\int_V {\mathbf A}\cdot{\mathbf B}dV
\label{eq:helicity}
\end{equation}
where $\mathbf{A}$ is the vector potential and ${\mathbf B}$ is the magnetic field such that ${\mathbf B}=\nabla\times{\mathbf A}$.  Observable signatures of this quality of the field are few, but \citet{Volegova:2010go} have demonstrated with simulations that helical turbulence would leave a statistical signature in the joint probability distribution of the polarization fraction, $p$, and the Faraday RM.  This effect is further discussed theoretically in \citet{Brandenburg:2014gp}, which illustrates how this asymmetry arises in a synchrotron-emitting region from the addition of the emission and the Faraday rotation when the field has non-zero helicity.  In brief, the Faraday rotation can either cause the emitted polarization vectors to wind up faster (and~therefore cancel more quickly) or more slowly (and therefore add more coherently) depending on the sign of the helicity.  This leads respectively to a negative (positive) shift in the PDF for a positive (negative) helicity.  (See Figure\,3 of \citet{Volegova:2010go}.)  The~resulting observable signature is summarized in Figure\,\ref{fig:cartoon}b, where because this observable depends on Faraday rotation, it depends on observing angle as illustrated.  Efforts are ongoing by West {et al., in prep} to use this method to look for helicity in the large-scale GMF.

An intriguing possibility for measuring the helicity of extragalactic magnetic fields through its signature on diffuse $\gamma$-ray emission is discussed in \citet{tashiro:2014} and references therein.  However, there is not yet a corresponding effect within the Galaxy.  It may become important to find such probes as our large-scale GMF models become more realistic (see Section~\ref{sec:beyond_adhoc}) and likewise our treatment of the turbulence at small scales (see Section~\ref{sec:turbulence}), since helicity is important to both contexts.

\subsection{Thermal Electrons---WHIM}

The warm/hot ionized medium (WHIM) plays several roles in GMF
studies.  Firstly, the Faraday RM discussed above depends on the free
electrons in this phase of the ISM.  Secondly, the WHIM emits thermal
Bremsstrahlung emission, a.k.a. free-free emission, from the radio to
microwave bands.  Its~spectrum ($\beta\simeq -2.1$) is not different enough from that of
synchrotron emission ($\beta\simeq -2.5$ to 3) for the components to
be reliably separated in the Galactic plane.  This has implications
discussed below in Section~\ref{sec:challenges}.

The most widely used model so far is the ``NE2001'' of
\citet{cordes:2002}, where the dispersion measures of Galactic pulsars
were used to fit a four-arm spiral model for the distribution of
thermal electrons.  The~model also includes a molecular ring around the inner Galaxy and local features in
several directions around the Sun.  The~vertical scale height was corrected in
\citet{gaensler:2008}, but~the model has otherwise remained the main
large-scale model of this component of the ISM.  Recently, a~new model
has appeared from \citet{Yao:2017kb} (YMW16).

One of the main issues with these models is the uncertainty over how
to handle the clumping of the WHIM and the fact that it may be
correlated (compression) or anti-correlated (pressure
equilibrium) with the small-scale fluctuations in the GMF
\cite{Beck:2003ke,sun:2008}.  Such correlations will introduce biases
in the inferred magnetic field strength, and there is no consensus on
how to handle this.

\subsection{Dust Grains}

The distribution of dust grains in the ISM can be probed by several observables, particularly its thermal emission in the sub-millimeter bands and its absorption of starlight.   The~difficulty here is that both the environment and the detailed properties of the dust grain populations vary throughout the Galaxy.  The~\planck\ observations of polarized dust emission have given us a new tool to study both the properties of the cold dust and its relationship with the magnetic field in the ISM.  See the review by Boulanger in this issue.  Recent results from the Gaia mission have given us much more detailed 3D information about the dust distribution in the local quadrant of the Galaxy by combining measurements of the dust extinction toward billions of stars with precise distance measures.  This can be used to map the local ISM in detail as done by \citet{Lallement:2018kl}.

\subsection{Galactic Cosmic Rays}\label{sec:crs}

The relativistic particles in the ISM that produce the synchrotron emission are thought to be generated by supernovae, but neither the origin nor the propagation of these cosmic rays is well understood.  The~higher energy particles (greater than a few GeV) produce inverse Compton emission that dominates the diffuse $\gamma$-ray sky, while the lower energy particles dominate the synchrotron emission spectrum in the radio to microwave bands.   \citet{Bernardo:2015ik}  and \citet{orlando:2018} show recent and complementary multi-wavelength analyses of these diffuse CRs with different propagation models.  These analyses can help modeling the synchrotron observables, and vice versa.  One of the more complicated unknowns is the question of anisotropic cosmic-ray diffusion.  Not only is the diffusion physics not well understood, but it determines the dependence of the CR propagation on the magnetic field.  Inferring the GMF from synchrotron emission therefore requires understanding the~diffusion.

It is worth mentioning that there are several CR propagation codes with different input physics being used to constrain the CR distribution with both the diffuse multi-wavelength emission data as well as the directly detected particle spectrum.  These include:  the GALPROP\footnote{\url{https://galprop.stanford.edu/}.} code of \citet{Strong:2009tq} that has been used for many years and most recently by \citet{orlando:2018}; the DRAGON code of \citet{Evoli:2008hz} used by other groups such as \citet{Bernardo:2015ik};  and the PICARD code of~\citet{Kissmann:2014hy}.  Many GMF modeling analyses separate the CR modeling from the field modeling (i.e., assume a fixed model for the former), but~ \cite{jaffe:2011,orlando:2013} show that they should be modeled together because of their interdependence.

\section{Models and Analyses}
\label{sec:models}

There have been several studies of the GMF over the past few
decades, and though there are some common features (e.g., spirals in disks), their morphologies
have a surprising variety in the details.  Some analyses include an exploration of
the parameter space and a quantification of statistical uncertainties,
but few have accounted for systematic uncertainties or have
quantitatively compared the different parametrized models that have be
explored independently.  These models can be roughly grouped into those that are
largely ad hoc constructions built to be compared with specific observations, and
those that arise from theoretical work.  Clearly, some physical properties
such as the divergence-free condition can be enforced on the ad hoc
models, and equally clearly, the observations inform the theoretical
work.  However, the two approaches can be considered to be complementary, and
if they do not meet up in the middle, they will each continue to be useful to compare.

The first half of this section reviews the analyses that have been done recently, with emphasis on their pros and cons compared to the others.  This collection of models contains many common morphological features, and since it is those physical features that are of interest, and how well they are constrained, we summarize them in the second half of this section.

\subsection{Current Magnetic Field Model Fits}

\begin{figure}[H]
\centering
$\begin{array}{ccc}

\textrm{(\textbf{a})}& \textrm{(\textbf{b})} & \textrm{(\textbf{c})}\\
 \includegraphics[width=45mm]{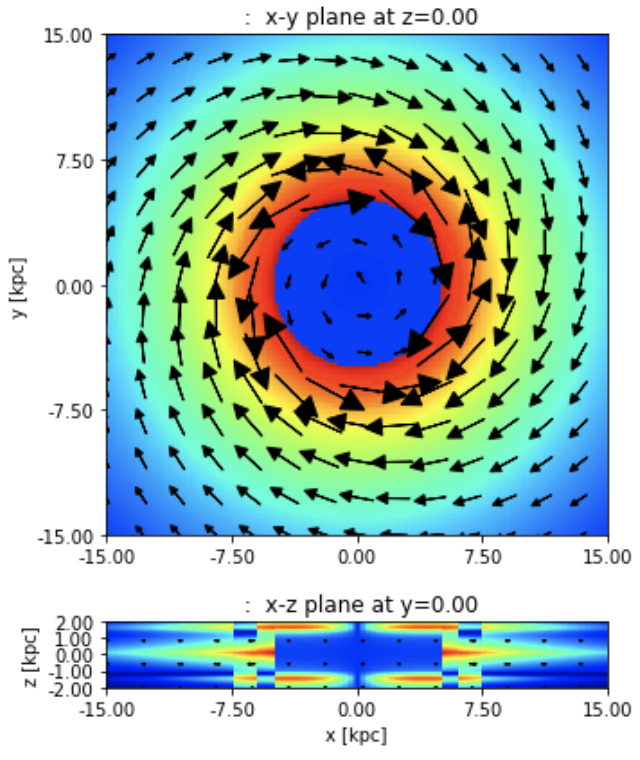} &
 \includegraphics[width=45mm]{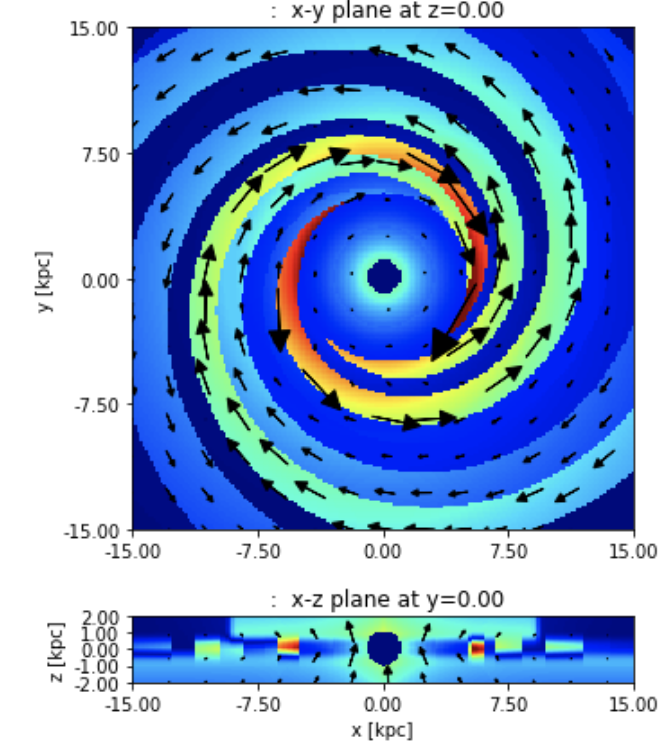} &
 \includegraphics[width=45mm]{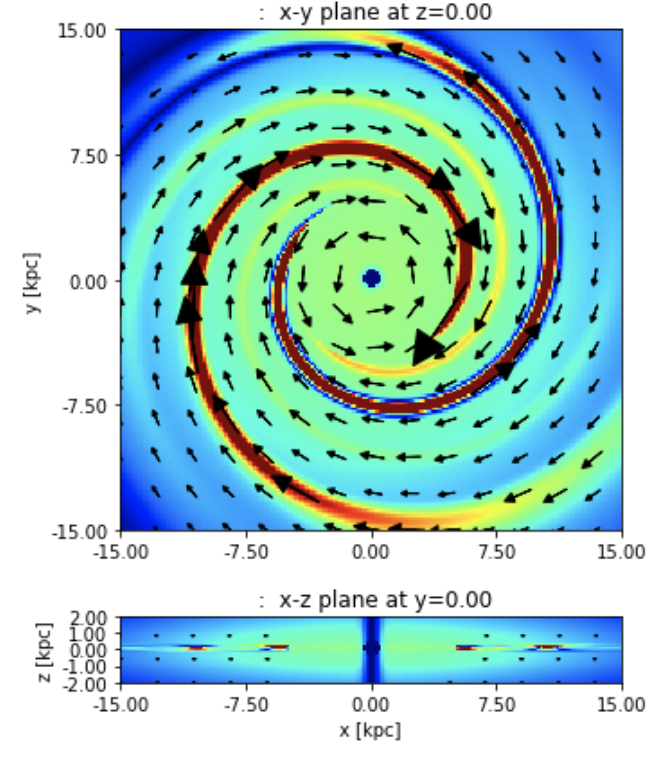} \\

\textrm{(\textbf{d})}& \textrm{(\textbf{e})} & \textrm{(\textbf{f}) }\\
 \includegraphics[width=45mm]{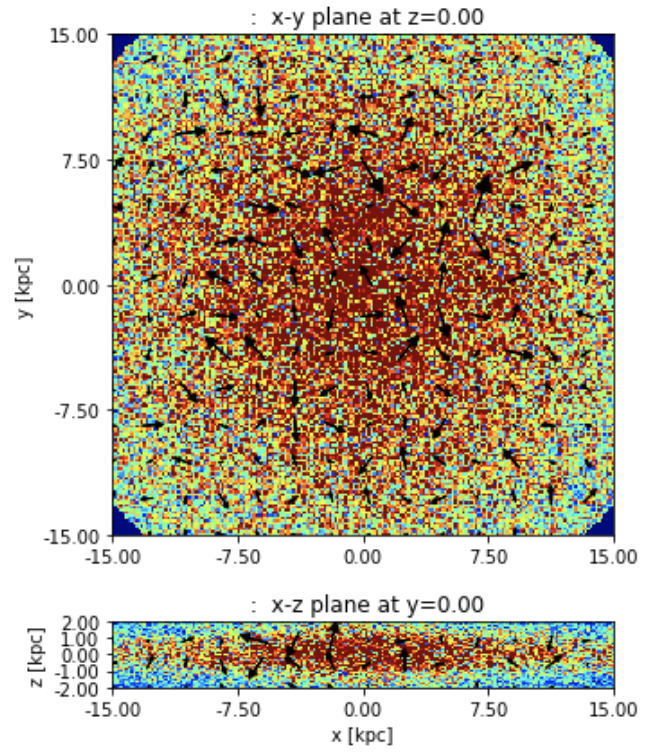} &
 \includegraphics[width=45mm]{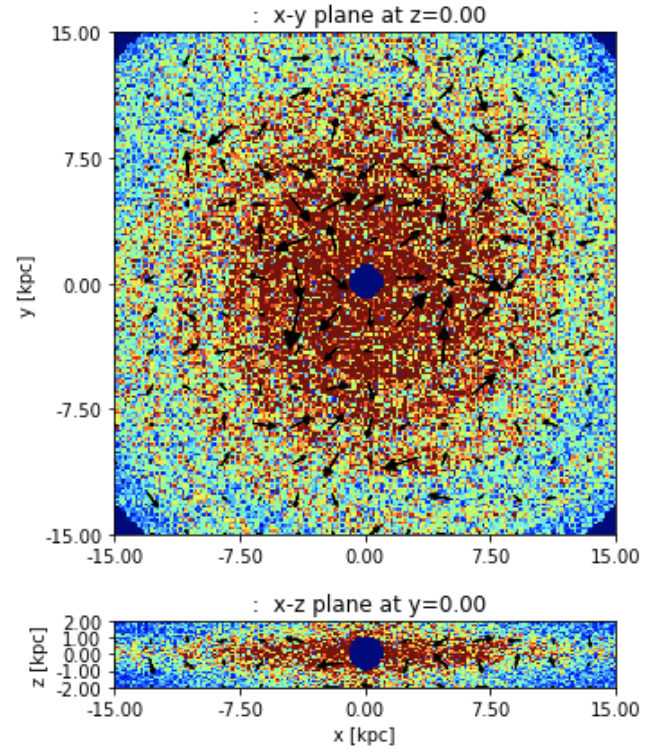} &
 \includegraphics[width=45mm]{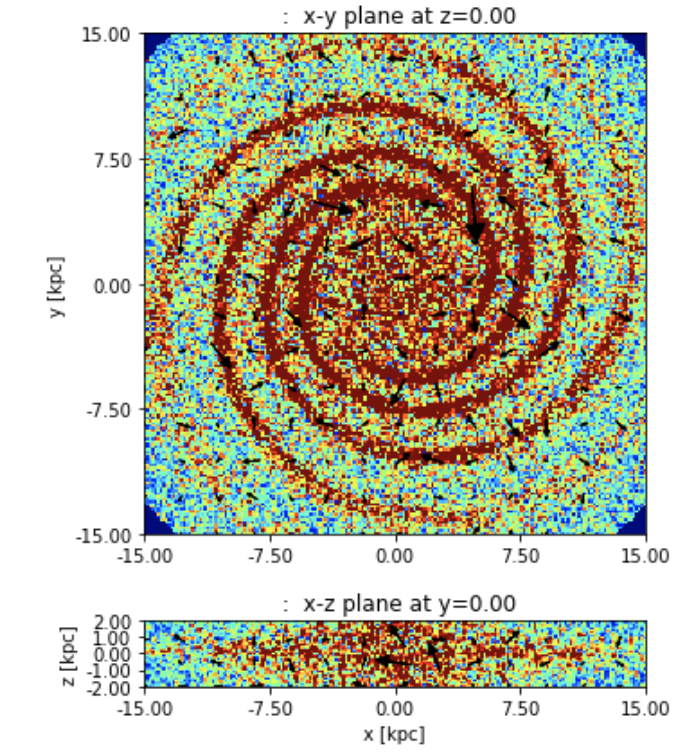} \\

\textrm{(\textbf{g})}& \textrm{(\textbf{h})} \\
 \includegraphics[width=45mm]{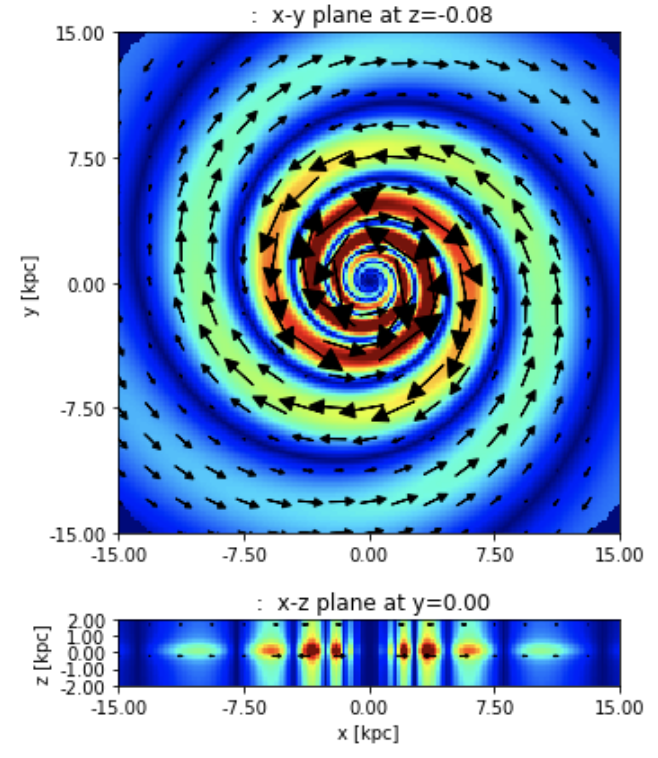} &
 \includegraphics[width=45mm]{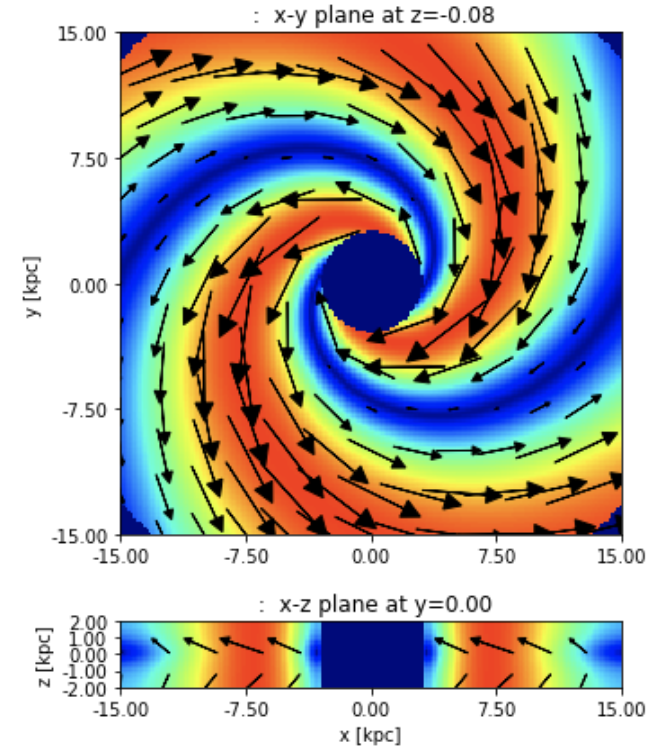} \\

\end{array}$
\caption{Examples of models for the GMF from the literature. (\textbf{a}) \citet{sun:2008} B$_\mathrm{coh}$; (\textbf{b}) \citet{jansson:2012c}  B$_\mathrm{coh}$; (\textbf{c}) \citet{jaffe:2013}  B$_\mathrm{coh}$; (\textbf{d}) \mbox{\citet{sun:2008}} B$_\mathrm{iso}$; (\textbf{e}) \citet{jansson:2012c}  B$_\mathrm{iso}$; (\textbf{f}) \mbox{\citet{jaffe:2013}}  B$_\mathrm{iso}$; (\textbf{g}) \citet{KACHELRIES:2007ho} B$_\mathrm{coh}$; (\textbf{h}) \citet{Fauvet:2011} B$_\mathrm{coh}$. (Models (\textbf{a}--\textbf{f}) have been slightly modified from their original parameters to match the \planck\ data \cite{pipxlii})  The~color represents the strength of the coherent magnetic component (arbitrary color scale), the arrows show its direction. The~top panel of each shows a cut through the Galactic plane at $z=0$ (the Sun is at $(x,y,z)=(-8.5 \,\mathrm{kpc},0,0)$), while the bottom panel of each shows a vertical cut intersecting the Sun and the Galactic center. }
\label{fig:views}
\end{figure}

This section presents an incomplete but representative sample of some of the current models in the literature, describing what datasets were used to constrain the models and what particular advantages or disadvantages each analysis had.  Some of these are shown in Figure\,\ref{fig:views} to illustrate the varieties of morphologies.  (Please note that this is a biased subset of only those that could be visualized in a consistent way using the modeling code {\tt hammurabi}\footnote{\url{https://sourceforge.net/p/hammurabicode/wiki/Home/}.}~\cite{waelkens:2009}.)

\begin{itemize}

\item  \citet{sun:2008} (refined in \cite{sun:2010}, ``Sun10'') first used the combination of synchrotron total and polarized intensity (at 408\,MHz and 23\,GHz respectively) along with RMs to compare several 3D models of the GMF.  They used the NE2001 model for thermal electrons and a simple exponential disk with a power law spectrum of index $p=-3$ for the cosmic rays.  This work included an analysis of the impact of the filling factor of the ionized gas in the ISM and examined several models from the literature, both axi-symmetric and bisymmetric spirals.  The~model they concluded was favored by the data was the ``ASS+RING'' based on an axisymmetric spiral disk with field reversals defined in several regions to match the data.  The~turbulence was modeled with a single-scale random field.  This was the first such analysis, though it was not  quantitative model fit, and it assumed a very high local cosmic-ray density to fit the data without an ordered random field component.

\item  \citet{jaffe:2010} (refined in \cite{jaffe:2011,jaffe:2013}, ``Jaffe13'') used these same synchrotron observables and the SGPS and CGPS extragalactic RMs to perform a systematic likelihood exploration in the plane of a 2D model based loosely on previous work by \cite{broadbent:1990}.  It used the NE2001 model for thermal electrons and a Galprop cosmic-ray model from \cite{strong:2010}.  It included an exponential disk to which is added four Gaussian-profiled spiral arms as well as a ring around the Galactic center.  This analysis first included realizations of the random components, both isotropic and ordered, based on a Kolmogorov-like GRF in an MCMC likelihood space optimization, but~only in 2D.  The~update in \citet{jaffe:2011} simultaneously constrained the CR lepton break at low energies in one of the first attempts to model the CRs and GMF simultaneously.  Then \cite{jaffe:2013} added the polarized dust information and saw how the different distributions of particles means that the two observables can perhaps constrain the GMF in different regions of the ISM.  These analyses, though, remain limited by the systematic uncertainties of the particle distributions.


\item  \citet{jansson:2012b} (refined in \cite{jansson:2012c}, ``JF12'') used the synchrotron total and polarized intensity from \wmap\ 23\,GHz as well as the 40k extragalactic RMs to perform a systematic likelihood exploration in 3D of a model with both thin and thick disk components, eight spiral arm or inter-arm segments, and an x-shaped halo field.  It was based on the NE2001 thermal electron density model (with the scale height correction from \cite{gaensler:2008}) and a CR model based on the ``71Xvarh7S'' from Galprop.  It used an analytical method to treat the random field components, and the measured pixel variance was used in the likelihood.  (See Section~\ref{sec:turbulence})  This was the first 3D model optimization with an MCMC likelihood analysis, but the use of the \wmap\ synchrotron map at 23\,GHz meant that the extra total intensity foregrounds biased their estimate of the random field component.   See also Unger \& Farrar below for updates.

\item \citet{han:2018} used both Galactic and extragalactic radio sources to model the RM reversals in the Galactic plane with a set of spiral arm and inter-arm segments.  The~analysis used the YMW16 model for thermal electrons.  This is not a global GMF model for the Galaxy, but an analysis specifically focused on where the field reversals lie using the distance information from pulsars.

\item \citet{Terral:2017bx} (``TF17'') used the spiraling x-shaped field models derived in \cite{ferriere:2014} to fit the RM data.  They explore both axisymmetric and bisymmetric possibilities.  This work represents the first quantitative fitting to models of spiraling x-shaped fields in theoretically derived forms (rather than ad hoc).

\item {Unger \& Farrar} \cite{Unger:2017ty} built on the JF12 work by replacing the ad hoc x-shaped halo field by the models of \citet{ferriere:2014}.  They also compared the results of fits based on different datasets (\wmap\ synchrotron total intensity versus 408\,MHz), different thermal electron models (NE2001 vs. YMW16), and CR distributions from the original work compared to those of \cite{strong:2010,orlando:2013}.

\item \citet{Shukurov:2018vb} derived eigenfunctions of the mean-field dynamo equation that can be used to construct any model consistent with those assumptions.  Though this analysis does not present one model fit to the data, it provides a framework for fitting more physically realistic models in future with a publicly available software package.

\end{itemize}

\subsection{Magnetic Field Morphological Features}\label{sec:model_features}

The rest of this section discusses the morphological features that are astrophysically interesting and are common among many if not all the different models, such as magnetic arms, reversals, and~x-shaped vertical fields.

\subsubsection{Axisymmetric Spirals}

Simple models began with axisymmetric spirals (e.g., \cite{page:2007,sun:2008}) with, e.g., exponential disks, and though one of these alone cannot reproduce all the observables, the morphology remains a component of many different models.  Sun10, for example, uses such a model as the basis on which reversals are added in an annular region and/or a spiral arm segment.  Jaffe13 adds spiral arms to an axisymmetric spiral base model.  JF12 uses multiple disk components to model the GMF in the thin and thick disks, where the spiral arm segments are imposed on the former.  Until the GMF can be modeled without parametrized ad hoc models (e.g., using the eigenfunctions of \citet{Shukurov:2018vb}), the basic axisymmetric spiral will remain useful.

The parameters of the axisymmetric spiral, however, are not yet well constrained because of the uncertainties in the equivalent parameters of the particle distributions.  In the case of CRs, for example, the thick disk scale height is not known even to within a factor of two \cite{orlando:2013}, and this translates into a corresponding uncertainty in the magnetic field strength as a function of height above the disk.  The~thermal electron density is better known, and so for example, the Sun10 model was corrected by a factor of two change in that model, but as discussed by those authors, there remains some uncertainty.  Recent analysis by \citet{Sobey:2019bz} of pulsar data from LOFAR\footnote{\url{http://www.lofar.org/}.}
estimate the GMF scale height assuming the \citet{Yao:2017kb} model for thermal electrons, but the paper also discusses how these systematic uncertainties may affect their estimates of both the scale height of the coherent magnetic field and its overall strength.
The pitch angle of the spiral is often assumed to be \hlo{$-11.5^\circ$} in the disk (Sun10, JF12, Jaffe13) following the NE2001 electron density model, and the RMs are consistent with this.  \citet{steininger:2018} showed that when allowed to vary, this pitch angle is not constrained by full-sky maps of RM and synchrotron emission, though this may be due to a pitch angle that has one value in the Galactic plane on average and another value in the local neighborhood that dominates the measures that are at higher latitudes.

Though the model fits discussed above provide numbers for these parameters, and some of them with small statistical error bars, these systematic uncertainties make it difficult to conclude that they are really constrained.

\subsubsection{Spiral Arms}

Though we have mapped out the spiral structure of our Galaxy's stellar component, it is harder to determine whether the GMF has a similar structure because of the difficulty determining distances to the corresponding measurements.  In several external galaxies that we can see face-on in synchrotron emission, the fields appear to be strongly ordered in spiral arm structures that may or may not be coincident with the material spiral arms.  Observations of these so-called magnetic arms are reviewed by \citet{beck:2013}.  The~question is of particular interest for what the magnetic arms may say about the mean-field dynamo, as discussed by \citet{Moss:2013tb}, for example.  For this reason, most models of the GMF include magnetic arms of some sort, whether as explicit bisymmetric spirals, ad hoc but continuously defined spiral arms \cite{jaffe:2013}, or as discontinuous segments~\cite{jansson:2012b,han:2018}.  If~the extragalactic RM features along the plane are indeed tracing these large-scale arms, then the models largely agree where the arms with the strongest coherent fields are.  However, since the models so far generally constrain the magnetic arms to follow the disk field pitch angle, the same systematic uncertainties above apply.

Comparison of the polarized emission of synchrotron and dust has the potential to probe whether the two components come from different spiral arm regions, as discussed by \citet{jaffe:2013}, but again, this depends on a better understanding of the field ordering.

\subsubsection{Reversals}

As reviewed in \citet{haverkorn:2014}, one of the most obvious features of the sky traced by RMs toward Galactic pulsars as well as extragalactic radio sources is the clear evidence of reversals in the large-scale GMF along the Galactic plane.  Though it remains difficult to determine the distance to these reversals, \citet{han:2018} use Galactic pulsars with distance estimates to model the reversals with alternating magnetic arms and inter-arm regions. \citet{Ordog:2017wc} recently added the analysis of the RM gradient in the diffuse polarized radio emission.  Their findings highlight the importance of modeling these reversals in 3D and connecting them to dynamo theory.

If indeed the GMF reversals are a feature of the large-scale Galactic structure and can be defined as reversals between magnetic arms, then we can perhaps learn about such structures from external galaxies.  See \citet{beck:2013} for a review of how observations of the RM of the diffuse emission can be combined with the rotational velocity information but are not yet sensitive enough to confirm the sorts of reversals we see in our own Milky Way.

The modeling projects discussed above include large-scale GMF reversals as either magnetic spiral arms or as annuli or both.  However, only \citet{han:2018} use the distance information from Galactic pulsars that can constrain where there are.  Models that use only extragalactic RMs and assume that the field along an entire magnetic arm is oriented the same way (Sun10, JF12, Jaffe13) do  agree on which segments are reversed.  However, the Han et al. analysis considers the direction in different sections of each arm, and these are not required, nor found empirically, to be uniform in each arm.  Increased pulsar sampling will help, as will adding information from other observables such as Zeeman splitting of masers \cite{Fish:2003gp,Han:2006hs,Green:2012fm}.

\subsubsection{Vertical (Poloidal) Field}

Early models fit to the GMF included no vertical component to the field, because starlight polarization observations showed clearly a field that remained parallel to the Galactic plane in the disk, and the strongest synchrotron emission in the plane likewise.  Observations of external galaxies seen edge-on in radio polarization initially showed a field largely parallel to the disk.  However, more sensitive radio observations of external galaxies where fainter emission from the ``halo'' or ``thick disk'' component could be observed showed an x-shaped magnetic field structure.  (Again, see \citet{beck:2013} for a review.)  Such a vertical component is again connected with the Galactic dynamo and winds.

The \citet{sun:2010} model included such a vertical component, as did \mbox{\citet{jansson:2012b}}, both cases of ad hoc x-shaped components.  \citet{ferriere:2014} more generally classified the different field morphologies that could have such a vertical component.  Both \mbox{\citet{jansson:2012b}} and the fitting of \citet{Terral:2017bx} conclude that such a component is likely favored by observations of our Milky Way.  However, again, the systematic uncertainties in the particle distributions keep this question open for the time being.

\subsubsection{Beyond the Ad Hoc}\label{sec:beyond_adhoc}

\citet{ferriere:2014} began the work of looking beyond ad hoc parametric models by using the Euler method to define convenient field configurations that could reproduce a spiral and an x-shaped vertical field.  Then next step was taken by \citet{Shukurov:2018vb}:  determining the eigenfunctions of the mean-field dynamo equation.  These functions can then reproduce any GMF that is physically possible within those assumptions.  Though parametrized models will always be useful for studying specific identified features of the large-scale GMF, we should increasingly move beyond them and exploit these more physical representations of the possible morphologies.  \citet{Terral:2017bx} fit the \citet{ferriere:2014} models to the RM data, and this work can now be combined with other tracers, as partly begun by \citet{Unger:2017ty}, who combined the \citet{ferriere:2014}  models with the ad hoc JF12 model.

\subsubsection{Turbulent Field}\label{sec:turbulence}

The treatment of the small-scale fluctuations in the GMF is one of the thornier questions that must be addressed in any modeling, even when the large-scale GMF is the only goal.  One reason is that small but local structures project to large angular-scale structures on the sky and can have a large effect on the model fitting if not properly taken into account.  Furthermore, there may be systematic correlations between fields and particles on small scales that must be accounted for in modeling the large scales, whether explicitly or statistically.  Lastly, when comparing models to data, it~must be quantified how far the latter are expected to deviate from the former.  The~Milky Way should be considered one realization of a galaxy model we are looking for, and that model includes some ``galactic variance'' due to the expected small-scale fluctuations that are model dependent.

The ISM is known to be turbulent at a range of scales (again, see \cite{haverkorn:2014} and references therein), and this turbulence is neither expected nor observed to be Gaussian.  Both properties present a challenge for generating simulated galaxy models and for comparing the data to the simulated observables.  Some modeling efforts simply ignore the stochasticity by fitting mean-field models to observables such as averaged RMs that depend only on that coherent field component.  This is effectively what Han et al. \cite{han:2018} do fitting RM vs distance plots for different regions of the Galaxy, and the error bars include the scatter that is partially due to the ISM turbulence.  The~JF12 model includes an analytic expression for the average contribution to each observable from the turbulence under a few simplifying assumptions.  This allows the average contribution to, e.g., synchrotron polarization to be correctly reproduced.  To compare that average to our Galaxy that itself is a single realization of a field with a random component, JF12 used the data essentially to bootstrap this statistically, so that the optimized model did take this measured variance into account in the likelihood.  However, as they point out and as further discussed in \cite{pipxlii}, the variance itself is an observable that should be used to improve the constraints on the degree of ordering in the magnetic field.

The next lowest order approximations are to create realizations of the random component during the simulation process by simply adding a randomly drawn number (usually from a Gaussian distribution), or a set of three for a random vector.  This can be done either to every pixel of a simulated sky map, to every point along a simulated LOS, or to every 3D voxel over which the simulated observables are integrated.  The~first approach is in a sense effectively similar to the JF12 method of simulating the ensemble average observed sky and comparing with a bootstrapped estimate of the variance.  The~second approach was used by O'Dea et al. \citet{ODea:2011bm} (with a further refinement discussed in the next paragraph), which then takes into account LOS averaging effects (i.e.,~depth depolarization) for each observed pixel.  The~third approach was used in the modeling of Sun10 and (effectively) in TF17, which then includes some effects of averaging within the observing angle (beam depolarization).

Adding information about the two-point correlation function of the turbulence is the next step, i.e., simulating a GRF with a given, e.g., Kolmogorov, power spectrum.  O'Dea et al. used such a prescription for generating the 1D turbulence for each LOS.  Jaffe13 used this in a 2D analysis restricted to the Galactic plane.  The~full 3D approach was used in the \planck\ analysis \cite{pipxlii}  (without any quantitative parameter optimization) and in \citet{steininger:2018} (with a full MCMC likelihood exploration).  This last result shows that it is now computationally feasible to do this.

The next step will be to include more information than simply the two-point statistics of the magnetic field.  Studies of the ISM turbulence have begun to characterize a variety of its statistical properties based on the diffuse synchrotron emission in total \cite{2017MNRAS.466.2272H} or polarized intensity \cite{Gaensler:2011ix,burkhart:2012,2018ApJ...855...29H}.  Likewise, for dust, the \planck\ collaboration has opened a new window into the turbulence in the colder phase of the ISM with high-resolution maps of the polarized dust emission (\cite{p18xii} and references therein).  These studies make use of MHD simulations with known physical parameters to study how to infer the physics from the observables.  Those simulations in turn can encode different assumptions about the turbulence and about the correlations among the relevant physical quantities such as the field strength and direction and the particle distributions (whether thermal electrons, CRs, or dust etc.).  See, e.g., \citet{Stepanov:2013ce} for comparisons of data and simulations focusing on the cosmic rays in MHD simulations, or \citet{pipXX} or \citet{Kandel:2017uu} for discussions of the dust.  With these studies, we can then use the information learned from MHD simulations to define physical parameters of the turbulence that the data may constrain.

\section{Challenges}\label{sec:challenges}

The observables available to us for studying the GMF are summarized in Section~\ref{sec:observables} and in Table\,\ref{tab:observables}, including some of their dependencies and drawbacks.  These issues are discussed in detail in the \planck\ paper on GMF modeling \citet{pipxlii} and summarized here.

\subsection{Synchrotron and CR spectra}

The degree to which synchrotron emission is polarized could be a direct tracer of the ordering of in the GMF, but only if we can compare the total and polarized emission components at the same frequency.  As discussed in Section~\ref{sec:obs_sync}, however, this is complicated by the presence of Faraday effects at the low frequencies and other emission components at the higher frequencies.  To then estimate the field ordering, we need to compare the data at different frequencies and therefore to understand the synchrotron emission spectrum, which in turn depends on the cosmic-ray energy~spectrum.

Multi-wavelength observations of the synchrotron emission can help us to understand the variations in its spectrum both across the sky and at different energies.  Studies such as \citet{kogut:2012} and \citet{fuskeland:2014} have quantified these variations based on available data, and \citet{pipxlii} shows how the variations impact the modeling that has been done so far.  In particular, the~synchrotron spectrum is $\beta\sim-3$ or steeper at high latitudes and high frequencies, but is harder at both low latitudes and low frequencies by of order $\Delta\beta\sim0.1$ or more.  This uncertainty in $\beta$ translates into a change in the synchrotron intensity of $\sim$50\% when comparing the total synchrotron intensity at 408\,MHz and the polarized intensity at 30\,GHz, with a corresponding impact on the estimate on the field ordering.

An independent and yet related way to make progress on this issue is through modeling of the diffuse $\gamma$-ray emission.  The~CRs at the relevant energies can be directly measured near the Earth, but their distribution is strongly affected by solar modulation and therefore is not representative of the ISM in general.  With additional observations from  {\it Voyager 1} \cite{Cummings:2016ks}, the interstellar spectrum in the solar neighborhood but unaffected by that modulation can now be used.  The~diffuse $\gamma$-ray emission in combination with the directly measured CR lepton spectrum can then be used to study the CR distribution.   See, e.g., \citet{orlando:2018} and references therein for an extensive analysis, including a warning that even with data that is unaffected by solar modulation, the spectrum may not be representative of the ISM on average.  Please note that the $\gamma$-ray emission in the relatively high-energy Fermi LAT bands is dominated by a different population of CRs than those that produce the synchrotron emission, so further progress may require a medium-energy project that measures the $\gamma$s in the few-MeV range.  \citet{orlando:2018} include predictions for next-generation medium-energy $\gamma$-ray observatories that will probe the energy range dominated by the inverse Compton emission of the same CR population.

All that can be confidently concluded at this point is that the anisotropic random field component (i.e., that which contributes to synchrotron polarization but not to Faraday RM) is of the same order of magnitude as the isotropic random component.  However, refining that estimate will require a re-analysis of the $\gamma$-ray, CRE, and synchrotron data including more intermediate frequencies.  With that additional data, we may be able to then trace the variation of the degree of field ordering, e.g., between and among the spiral arms (e.g., as \citet{haverkorn:2004} did with RMs on smaller scales), which will tell us about the relationship between the large-scale Galaxy dynamics and the interstellar turbulence.

\subsection{Local Features}
\label{sec:local_features}

Another challenge in studying the large-scale structure of the GMF is distinguishing morphological features that are large-scale on the sky because they are close by from those that are physically large-scale components of the Galaxy.  The~NPS is the most obvious example of how such a large angular-scale feature may impact GMF or CMB studies \cite{Liu:2014td}, but it is not the only one.

\subsubsection{Loops and Spurs}

The ISM turbulence discussed in Section~\ref{sec:turbulence} is thought to be injected by supernova remnants that expand into the ambient ISM, compress gas and magnetic fields, and start a cascade of turbulence from the scale of the SNR ($\sim$100\,pc) down to smaller scales following a Kolmogorov-like power law.  That discussion of the turbulence refers to the smaller scales where the morphology of the SNR is no longer relevant.  However, the SNRs themselves are a significant part of the ISM that cannot be treated either as the large-scale GMF or as the small-scale turbulence.  One approach is to mask out regions of the sky that are thought to be dominated by a local object such as the NPS (as done in most of the modeling discussed above).

The NPS looked at by \citet{Liu:2014td} is only the largest of the loops and spurs identified in radio data for some decades.  Mertsch \& Sarkar \citet{Mertsch:2013gg} studied the impact on the radio sky of a collection of synchrotron-emitting shells distributed roughly as we expect supernovae to be distributed in the Galactic disk.  The~observed power spectrum of the synchrotron emission at 408\,MHz cannot be explained by a combination of a large-scale disk field and small-scale Gaussian turbulence, which is how it is often modeled.  Mertsch \& Sarkar showed that it can be better reproduced by the addition of such correlated but turbulent structures as loops and spurs at a variety of scales, most of which cannot be seen by eye in the maps.  More recently, \citet{vidal:2015} have studied in detail the numerous loops and spurs visible in the \planck\ maps in polarization and conclude that even when not bright enough to be obvious by eye, these features will perturb both studies of the GMF as well as the separation of the synchrotron foregrounds from the CMB.

It remains to be quantified how such features can affect GMF model fits.  For example, the NPS visibly extends nearly from the Galactic plane to the north pole and has an emission ridge that lies near the zero-longitude meridian.  This means that the observed anti-symmetry of RMs across the plane toward the inner Galaxy (the butterfly pattern discussed in \cite{1997A&A...322...98H}) may be affected by this structure \cite{Wolleben:2010bl}.  Understanding the impact of these loops and spurs on the observables is therefore necessary to fitting global models to the GMF and interpreting them physically.

\subsubsection{Fan Region}

In the outer Galaxy, the most obvious feature of the sky in the radio is the region of polarized emission in the second quadrant extending around the Galactic plane and up to middle latitudes ($b\lesssim30^\circ$).  This region is known as the Fan region due to the appearance of radio polarization maps, where the orientations of the polarization vectors ``fan'' outward from the plane due to Faraday rotation.  At higher frequencies where Faraday effects are negligible, this region is very highly polarized by an inferred GMF almost entirely parallel to the Galactic plane, even well off the plane.  The~distance to this emission region is highly uncertain, and it remains unknown whether it is a local feature or a result of our view of the large-scale GMF in the outer Galaxy.  It is highly polarized in dust emission at 353\,GHz as well, and difficult to reproduce with large-scale GMF models \cite{pipxlii}.

\citet{hill:2017} have compared more recent Global Magneto-Ionic Medium Survey (GMIMS) data with other tracers such as H$\alpha$ and examined depolarization features that can be associated with regions of known distances.  They argue that a large fraction of this emission must be coming from a significant distance of $d\gtrsim2$\,kpc and that its asymmetric distribution about the plane (with significantly more emission above) can be explained by large-scale Galactic warp.  

Again, it is a better sampling of GMF tracers for which we have distance information (e.g., pulsars and starlight polarization) that is necessary to understand this region and its impact on large-scale GMF fits.

\subsubsection{Local Bubble}

Another local feature that affects the observations on the largest scales is the impact of what immediately surrounds our own solar system. A variety of studies have probed the local ISM with starlight polarization, dust reddening, Faraday tomography, etc.  See \citet{frisch:2011} for a review.  Recent data from the Gaia mission to map the distances to billions of stars constitute a significant improvement.  \citet{Lallement:2018kl} use the dust extinction combined with the Gaia parallax distances to map the nearby dust in 3D.  \citet{Alves:2018kw} have demonstrated how the polarization of dust emission can be used to study the local magnetic field.  When combined, these analyses have the potential to greatly improve our local field modeling, which will then in turn allow us to better constrain the large-scale GMF.

\subsection{Galactic Center, Outflow, and Fermi Bubbles}

Though not local, there are other features of the multi-wavelength sky that can impact the fitting of large-scale GMF models.  In 2010, the Fermi mission discovered lobes of $\gamma$-ray emission extending to $\sim50^\circ$ above and below the Galactic plane toward the Galactic center, since referred to as the Fermi bubbles \cite{Su:2010vk}.  These have been interpreted as evidence of giant Galactic-scale outflows.   \mbox{\citet{carretti:2013}} have connected these $\gamma$-ray features with spurs and loops of polarized radio emission seen in the S-Band All Sky Survey (S-PASS) data.

As with the NPS, it is difficult to establish distances to these features of the radio or $\gamma$-ray sky.  But~an outflow of some sort from the Galactic disk is the likely explanation for the vertical field component, the x-shaped morphology seen in other galaxies and perhaps our own \cite{jansson:2012b, Terral:2017bx}.  It is, therefore, an outstanding question whether modeling the large-scale GMF must then also include a modeling of the Fermi bubbles and the possible associated synchrotron emission in the radio and microwave bands.

Even in the plane, the central few kpc of the Galaxy constitute a key region we know little about.  Most of the above models assume a simple azimuthal field in the central region, or continue the ever-tightening spiral.  Approaching the center itself, some models simply set the field to zero, and none of those discussed above include a physical connection to any outflow or x-shaped halo field.  More physically motivated formalisms (e.g., \citet{ferriere:2014} or \citet{Shukurov:2018vb}) are needed to do this better.

\subsection{Sub-Grid Modeling}

Section~\ref{sec:turbulence} discussed the complicated question of the turbulence in the magnetized ISM and what we can learn from high-resolution MHD simulations.  However, any finite simulation of the large-scale GMF will have a finite resolution below which the GMF cannot be modeled in detail, and this is well above the resolution at which interesting things happen in the MHD simulations.  Below this scale, the~modeling must encode our knowledge of how the smallest-scale fields interact with the matter on average and to use this ``sub-grid modeling'' to represent the scales we cannot probe.  Until we can run a full galaxy MHD simulation from kpc to sub-parsec scales, we can only use these simulations~indirectly.

With the notable exception of \citet{sun:2008}, most modeling assumes uniform magnetic field and particle properties over the smallest resolved grid cell, which for large-scale field modeling can be tens of parsecs.  For Faraday rotation, for example, this means assuming a smooth average thermal electron density and computing the RM with no account of the known clumpy nature of the ISM.  Any~small-scale (anti)correlation of the magnetic field with the particle distributions can result in an over- or under-estimation of the large-scale magnetic field strength.  Specific correlations with the magnetic field that are not taken into account in most of the above modeling studies include:  the ionized gas as discussed in \cite{sun:2008}, where a filling factor for the ionized gas was considered and its implications for the estimate of the magnetic field strength;  the relativistic cosmic-ray leptons discussed in \cite{Seta:2017id};  the dust grain density \cite{Kandel:2017uu} or alignment efficiency \cite{Fanciullo:2017cx}.  A better understanding of these correlations can be encoded in the modeling, or at very least the effects ought to be quantified so that we can estimate their impact on the large-scale GMF fits.

\section{Prospects}\label{sec:prospects}

As always, the way forward certainly includes continuing to gather more of the traditional observational tracers, e.g., more pulsar RMs and distances, more starlight polarization measurements and distances, more synchrotron frequencies between the radio and microwave bands, etc.   Gaia has already demonstrated how orders of magnitude more stellar distance and extinction measurements can impact our understanding of the local ISM \cite{Lallement:2018kl}.  Combining this with dust polarization measurements from \planck\ and additional starlight polarization surveys will likewise help us to map the magnetic fields in the solar neighborhood.  Next-generation radio surveys are poised to do the same for Galactic pulsars and extragalactic radio sources.  The~GALFACTS\footnote{\url{https://www.ucalgary.ca/ras/GALFACTS}.} team have completed their survey to improve our sampling of radio sources by an order of magnitude over the full sky visible from Arecibo.  The~LOFAR project is now underway to demonstrate the power of not only the idea of linking large numbers of radio antennae in a software telescope but also of opening the low-frequency window onto the Universe.  The~latter will allow us to probe the more tenuous regions of the ISM.  This is one of the first SKA-pathfinder projects testing technology and algorithms for the planned SKA that is expected to detect every pulsar in our Galaxy (that sweeps in our direction \cite{Keane:2014wk}) as well as orders of magnitude more extragalactic sources.  This improved sampling and distance information will help us to isolate the field reversals that we see in the RMs as well as the local features such as bubbles and loops.  A new survey of OH Masers \cite{Green:2012fm} will increase our sampling of Zeeman splitting measurements so that it can also tighten constraints on the large-scale field.

Constraints on Galactic cosmic rays will also be improved by additional direct measurements of the local interstellar spectrum by Voyager \cite{Cummings:2016ks} and future constraints in the medium-energy $\gamma$-rays by, e.g., the proposed AMEGO\footnote{\url{https://asd.gsfc.nasa.gov/amego/index.html}.} mission.  When combined with additional synchrotron maps at intermediate (a few GHz) frequencies such as from S-PASS, C-BASS, and QUIJOTE~\footnote{\url{http://www.iac.es/proyecto/cmb/pages/en/quijote-cmb-experiment.php}.}, we should be able to constrain the CR and synchrotron spectra at large scales and therefore the magnetic field ordering, at least on average in the disk.  If we then combine this information with the variations in the RMs from the vastly improved sampling from the SKA pathfinders, we may be able to start studying its variations in different regions of the Galaxy in 3D.

In addition to more data from the traditional tracers, we also have the prospect of new tracers and methods.  UHECRs are also deflected in the magnetic field, and though we cannot back-trace the particles to their sources, the anisotropy in the distribution of their arrival directions is a statistical probe of the local magnetic field (where the meaning of ``local'' depends on the particle rigidity). 
 \citet{hawc:2018} have recently measured the direction of the local interstellar magnetic field from the anisotropy in the HAWC and IceCube data at 10\,TeV.  Multi-messenger astronomy has the potential to identify the sources of these particles by associating them with photons, neutrinos, and gravitational wave events unaffected by the GMF, so~the individual UHECR deflections could be used as an additional tracer of the GMF.  New techniques include analyses of synchrotron data in new ways, such as the family of tools based on the polarization gradient of \citet{Lazarian:2018bi}.  See \cite{Ho:2019vh} for how these methods compare to Faraday tomography, for example.  This review has not discussed Zeeman splitting, because generally the number of measurements is not sufficient for probing the large-scale GMF.  However, these data can provide crucial independent probes in the regions where we do have them, and the sampling is soon to be improved by new projects such as MAGMO by \citet{Green:2012br}.  Though this is not a new magnetic field tracer, using a large sampling of measurements to trace the GMF may indeed soon have a new discriminatory~power.

However, in making progress on understanding the GMF, more data may not be more important than putting together information from different fields and from asking different questions.   The~IMAGINE project \cite{imagine} aims not merely to add observables such as UHECRs to the mix but more importantly to provide a common framework for analysis.  Just as a decade ago, significant progress was made by several teams combining three complementary observables, so we can expect the next advances to come from a yet more holistic approach.  Initial attempts have already been made to fit the CR spectrum simultaneously with the large-scale GMF \cite{jaffe:2011} and to add the dust polarization \cite{jaffe:2013}, but~it was not computationally feasible with the tools available at the time to go beyond a very simple analysis with too many assumptions.  \citet{steininger:2018} have published a more efficient platform into which we can plug in the information from the many disparate tracers and explore the likelihood space in a rigorous Bayesian analysis.  It is also important to move beyond the ad hoc models and simple field components described in Section~\ref{sec:B_components} to include higher-order moments of the random components as well as properties such as helicity.

It is also worth mentioning that simulations of Milky Way-like galaxies are becoming more informative about the amplification of magnetic fields during galaxy formation.  \citet{Pakmor:2017ia} for example show how the Auriga simulations reproduce spiral galaxies with magnetic fields with similar exponential disks to what we observe in our Galaxy and with similar strengths.  Whether these simulations will inform GMF modeling work or vice versa is an interesting question.

The bottom line for the time being is that modeling the GMF has now reached an interesting and challenging point, where there are both degeneracies and contradictions in the parameter space, but~still too many systematic uncertainties to know what to make of them yet.  However, the uncertainties are being attacked in several ways, with new techniques, new observables, and new combinations of old observables, and the prospects are correspondingly bright for improving our models.

\vspace{6pt}


\acknowledgments{T.R.J. thanks J. West for advice on the helicity cartoon in Figure\,\ref{fig:cartoon}.
}

\funding{\hlo{This research received no external funding}}

\conflictofinterests{The author declares no conflict of interest.}

\abbreviations{The following abbreviations are used in this manuscript:\\

\noindent
\begin{tabular}{@{}ll}
GMF & Galactic magnetic field\\
GRF & Gaussian random field \\
RM & Faraday rotation measure\\
CMB & cosmic microwave background\\
ISM & interstellar medium\\
MCMC & Markov chain Monte Carlo\\
MHD & magnetohydrodynamics\\
CR & cosmic ray\\
UHECR & ultra-high-energy cosmic ray
\end{tabular}}

\reftitle{References}


\end{document}